# Modified Froelich's Equation for Modelling of a Three Phase Self-Excited Synchronous Generator

Udayan Banerjee [1], Nayan Kumar[1], Thotakura NSC Sekhar [2], Tapas Kumar Saha [1], Chayan Kumar Paul[3]

[1] Department of Electrical Engineering, NIT Durgapur, India.
[2] School of Electrical Engineering, KIIT University, Bhubaneswar, India.
[3] Department of Electrical Engineering, IIEST Shibpur,India.
udayanbanerjee15@gmail.com

**Abstract*:* With advancement in design and analysis of electromechanical and electromagnetic devices, the modelling of magnetic saturation of a synchronous generator has emerged to be a subject of interest in number of publications. Most of the existing electrical machine modelling methods does ignore the saturation effect for simplicity. On the other hand, who incorporate saturation effect, are dealing with complex computation of coefficients which involves tedious curve fitting techniques like non-linear regression, least-squares. This paper presents the novel method of modelling of the self-excited synchronous generator along with magnetizing characteristics with ease and good accuracy which is inspired from Froelich's equation. The proposed mathematical model is implemented in simulation environment and validated the results with a practical three phase self-excited synchronous generator in which saturation plays a vital role.**

## 1. Introduction

The distributed electric utility is developed on the premise of decentralized power generation and close to intake sites. An enhanced approach to use the developing rising capability of distributed generation (DG) is to consider the power production and its related loads as a system called "microgrid". For these DGs to produce power, synchronous generators are one of the exceptional possible choices and will rule the power generation in future due to expanding share of distributed generation sources, furthermore for power production through renewable energy resources. The possibility of using a permanent magnet synchronous machine as a generator in stand-alone mode has been known from last century [1]. There are many attractive features of a permanent magnet synchronous generator (PMSG) over a conventional wound field synchronous generator (SG) such as brushless construction, higher efficiency, less maintenance and high torque/weight ratio. However due to presence of permanent magnet of its rotor, stator terminal voltage regulation is poor [2].

Louis, Schenectady & Frederick in 1946 patented self-excitation and regulating system for alternating current generator. By connecting a three-phase rectified connected from stator and supplied to the field winding through the regulatory system which consists of high reactance current transformer with saturable core connected at the stator terminals and whose secondary terminals are connected in delta and which is then fed to rectifier circuit [3]. Alger in 1964, modified the alternator design with magnetically hard steel alloy pole laminations 10% of total laminations to retain a good amount residual magnetism which helps in self-excitation and concluded that, this configuration can help building a small synchronous generator for isolated power supply that can supply to all electrical loads. The machine described doesn't require any external supply to excite the rotor but it does have slip rings and brushes later on self-excited brushless alternators both single phase and three phases were used [4]. In 1983, Shibata and Kohrin [5] proposed a new self-excited design, self-excited synchronous generators in conformity with the fundamental conception of the invention, constructions of such generators become simple. Various methods for implementing saturation have also been studied and compared [6]. But they are relatively complex to incorporate in order to realize self-excitation model.

This paper focuses on development and validation of a mathematical model for a Self-excited Synchronous Generator configuration as shown in Fig. 1, using Froelich's equation. The configuration involves a diode bridge feeding the generators field winding with a residual magnetism.

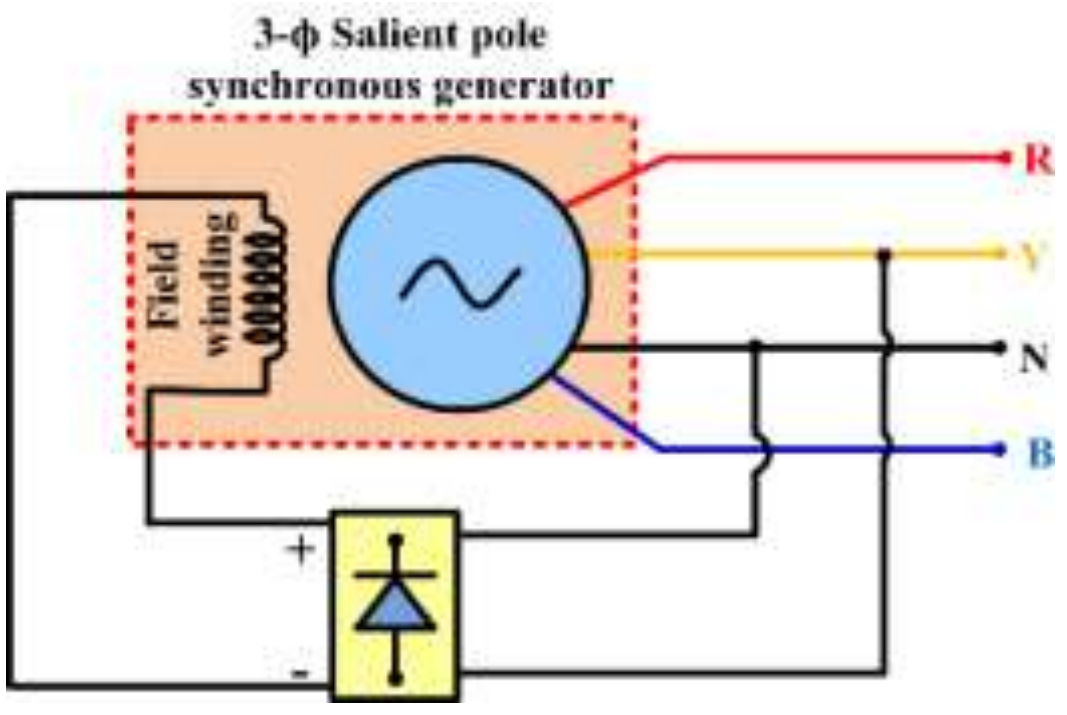

**Fig. 1.** *Schematic Diagram of SESG*

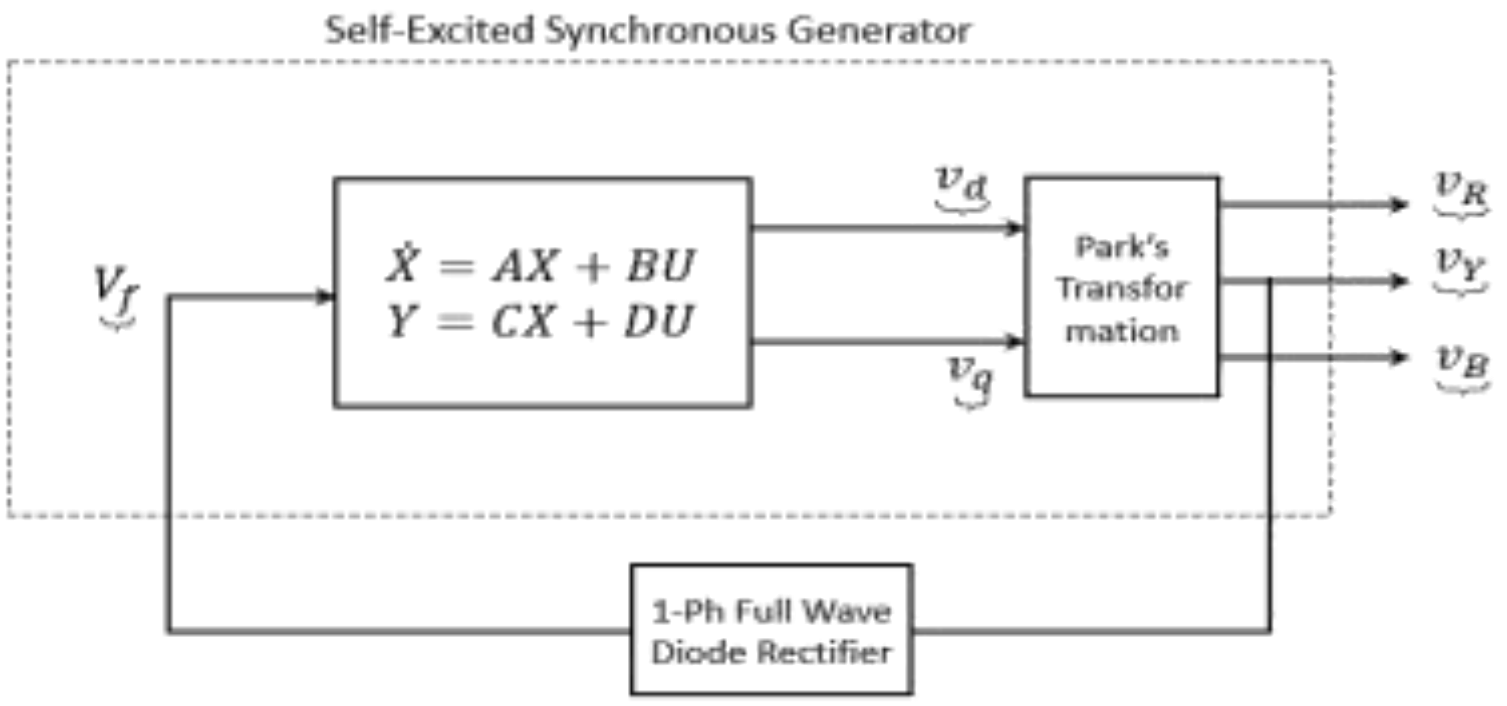

**Fig. 2.** *State Space based Schematic Diagram of SESG in d-q reference frame*

## 2. Synchronous Machine Modelling

Different mathematical modelling techniques exist to represent the transient behaviour of the synchronous generators. But they differ in the level of accuracy and complexity. The majority of them depend on well-known two-axis theory and are solved through digital computers. In any case modelling isn't always a truth, it's a method to get few inferences about the plant, subsequently, to reduce complexity in modelling a few assumptions were taken which are:

- Leakage inductances are impartial to the saturation, saturation affects the mutual inductances only.
- Saturation relationship among the resultant air-gap flux and the MMF while machine is under loaded conditions is similar to the no-load conditions.
- There is no any magnetic coupling between the d-and q-axes due to nonlinearities brought with the aid of saturation.
- Sinusoidal distribution of MMF in air gap of the machine.
- Saliency of rotor is taken into account and that due to stator slots is neglected.
- A lumped circuit model has been considered.

### 2.1. Modelling in a-b-c reference frame

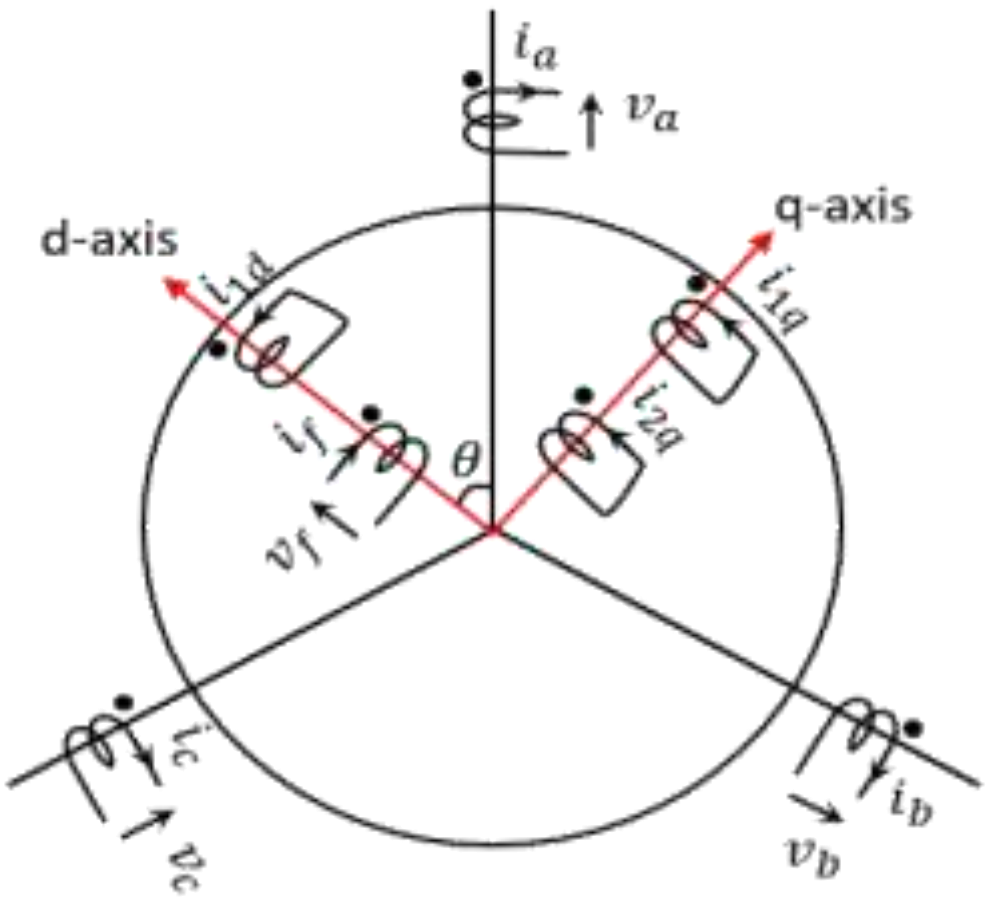

**Fig. 3.** *Schematic representation of Synchronous Generator with three damper windings in a-b-c reference frame*

Literature suggests that using three damper windings for round rotor and two damper windings for salient rotor can give relatively good accuracy in the response with reduced complexity[7] [9]. The model in this paper also considers three damper windings for the salient rotor synchronous generator, it can be easily extended to round rotor also. The equivalent schematic representation is shown in Fig .3. Equations (1) to (7) indicate the winding equations for the representation used in Fig. 3. [7] [9]. Stator and rotor flux linkage equations are shown in (7.1).

$$v_a = -R_a i_a - p\psi_a \tag{1}$$

$$v_b = -R_a i_b - p\psi_b \tag{2}$$

$$v_c = -R_a i_c - p\psi_c \tag{3}$$

$$v_f = R_f i_f + p\psi_f \tag{4}$$

$$v_{1d} = R_{1d} i_{1d} + p\psi_{1d} \tag{5}$$

$$v_{1q} = R_{1q} i_{1q} + p\psi_{1q} \tag{6}$$

$$v_{2q} = R_{2q} i_{2q} + p\psi_{2q} \tag{7}$$

$$\begin{bmatrix}\psi_s^{3x1}\\ \psi_r^{4x1}\end{bmatrix} = \begin{bmatrix}L_{ss}^{3x3} & L_{sr}^{3x4}\\ L_{rs}^{4x3} & L_{rr}^{4x4}\end{bmatrix}\begin{bmatrix}i_s^{3x1}\\ i_r^{4x1}\end{bmatrix} \tag{7.1}$$

$$\text{Where}[x_s] = \begin{bmatrix}x_a\\ x_b\\ x_c\end{bmatrix};\ [x_r] = \begin{bmatrix}x_f\\ x_{1d}\\ x_{1q}\\ x_{2q}\end{bmatrix}; x = \psi \text{ or } i$$

$$[L_{ss}] = \begin{bmatrix}L_a & M_{ab} & M_{ac}\\ M_{ab} & L_b & M_{bc}\\ M_{ca} & M_{cb} & L_c\end{bmatrix};$$

$$[L_{sr}] = \begin{bmatrix}M_{af} & M_{a1d} & M_{a1q} & M_{a2q}\\ M_{bf} & M_{b1d} & M_{b1q} & M_{b2q}\\ M_{cf} & M_{c1d} & M_{c1q} & M_{c2q}\end{bmatrix};$$

$$[L_{rs}] = \begin{bmatrix}M_{fa} & M_{fb} & M_{fc}\\ M_{1da} & M_{1db} & M_{1dc}\\ M_{1qa} & M_{1qb} & M_{1qc}\\ M_{2qa} & M_{2qb} & M_{2qc}\end{bmatrix};$$

$$[L_{rr}] = \begin{bmatrix}L_f & M_{f1d} & M_{f1q} & M_{f2q}\\ M_{1df} & L_{1d} & M_{1d1q} & M_{1d2q}\\ M_{1qf} & M_{1q1d} & L_{1q} & M_{1q2q}\\ M_{2qf} & M_{2q1d} & M_{2q1q} & L_{2q}\end{bmatrix}$$

### 2.1 Modelling in d-q reference frame

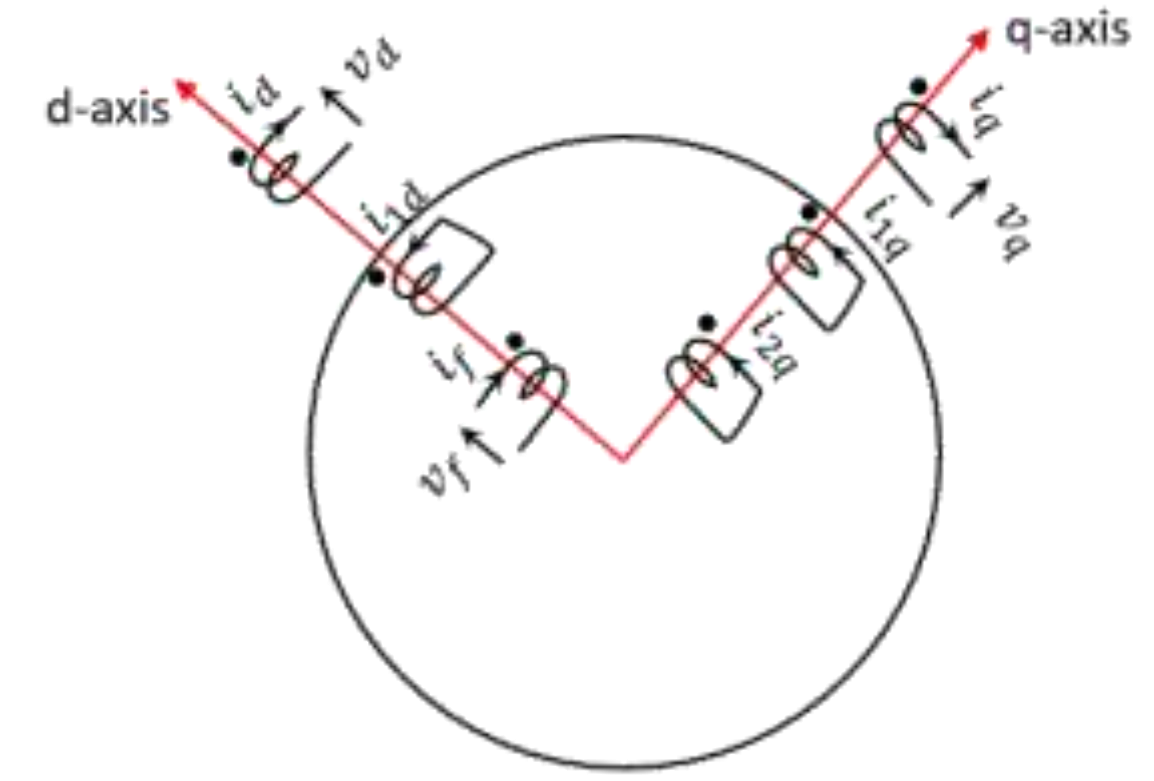

**Fig. 4.** *Schematic representation of Synchronous Generator with three damper windings in d-q reference frame*

$$v_d = -R_a i_d - \omega\psi_q - p\psi_d \quad (8)$$

$$v_q = -R_a i_q + \omega\psi_d - p\psi_q \quad (9)$$

$$v_0 = -R_a i_0 - p\psi_0 \quad (10)$$

$$v_f = R_f i_f + p\psi_f \quad (11)$$

$$0 = R_{1d} i_{1d} + p\psi_{1d} \quad (12)$$

$$0 = R_{1q} i_{1q} + p\psi_{1q} \quad (13)$$

$$0 = R_{2q} i_{2q} + p\psi_{2q} \quad (14)$$

$$\begin{bmatrix}\psi_d\\ \psi_q\\ \psi_0\\ \psi_f\\ \psi_{1d}\\ \psi_{1q}\\ \psi_{2q}\end{bmatrix} = \begin{bmatrix} L_d & 0 & 0 & M_{df} & M_{d1d} & 0 & 0\\ 0 & L_q & 0 & 0 & 0 & M_{q1q} & M_{q2q}\\ 0 & 0 & L_0 & 0 & 0 & 0 & 0\\ M_{fd} & 0 & 0 & L_f & M_{f1d} & 0 & 0\\ M_{d1d} & 0 & 0 & M_{f1d} & L_{1d} & 0 & 0\\ 0 & M_{q1q} & 0 & 0 & 0 & L_{1q} & M_{1q2q}\\ 0 & M_{q2q} & 0 & 0 & 0 & M_{1q2q} & L_{2q}\end{bmatrix}\begin{bmatrix} i_d\\ i_q\\ i_0\\ i_f\\ i_{1d}\\ i_{1q}\\ i_{2q}\end{bmatrix}$$

Fig. 4 shows the schematic of Synchronous generator in d-q reference frame and the governing winding equations are given in (8) to (14). MATLAB/Simulink platform is used to simulate the state space based self-excited synchronous machine model as indicated in Fig.2. In this, state equations are considered which explains the electrical dynamics with constant mechanical speed. From the above equations it is clearly understood that the following machine parameters are required to implement the model $R_a, L_d, L_f, L_{1d}, R_f, R_{1d}, M_{df}, M_{d1d}, M_{f1d}, L_q, L_{1q}, L_{2q}, R_{1q}, R_{2q}, M_{q1q}, M_{q2q}, M_{1q2q}$ these are to be provided by the manufacturers however, majority of them do not provide such information. Resistances $R_a, R_f$ can be measured by direct tests but not the damper bars resistances. Similarly, other inductances are also difficult to obtain from the direct testing unlike $R_a, R_f$.

Nonetheless, many researchers have proposed distinct ways in estimating these parameters which involve complex and tedious mathematical analysis [8].To reduce the analysis it is found in the literature to change the model shown in above equations in terms of following standard parameters $L_d, T'_d, T''_d, T'_{d0}, T''_{d0}, L_q, T'_q, T''_q, T'_{q0}, T''_{q0}$ ,to obtain these parameters a frequency response test needs to be performed [9]. But it is observed that finding these parameters are also complex and involves costly equipment. In alternative it is proposed to back calculate these values from few simple tests along with some simple calculations. Table 1 provides a comparison of those parameters. Equations (15) to (28) provide the relationships between the obtained test parameters and the required model parameters.

**Table 1** Modelling Parameters

| **Required parameters** | **Standard parameters** | **Parameters from simple tests** |
|---|---|---|
| $R_a, L_d, L_f, L_{1d}, R_f, R_{1d}, M_{df}, M_{d1d}, M_{f1d}$ | $L_d, T'_d, T''_d, T'_{d0}, T''_{d0}$ | $R_a, R_f, X_{d_{unsat}}, X_{q_{unsat}}, X_{ls}, X'_d, X''_d, X'_q, X''_q$ |
| $L_q, L_{1q}, L_{2q}, R_{1q}, R_{2q}, M_{q1q}, M_{q2q},$ | $L_q, T'_q, T''_q, T'_{q0}, T''_{q0}$ | |

$$X_d = X_{md} + X_{ls} \quad (15)$$

$$X'_d = \left(\frac{1}{X_{md}} + \frac{1}{X_f}\right)^{-1} + X_{ls} \quad (16)$$

$$X''_d = \left(\frac{1}{X_{md}} + \frac{1}{X_f} + \frac{1}{X_{1d}}\right)^{-1} + X_{ls} \quad (17)$$

$$X_q = X_{mq} + X_{ls} \quad (18)$$

$$X'_q = \left(\frac{1}{X_{mq}} + \frac{1}{X_{1q}}\right)^{-1} + X_{ls} \quad (19)$$

$$X''_q = \left(\frac{1}{X_{mq}} + \frac{1}{X_{1q}} + \frac{1}{X_{2q}}\right)^{-1} + X_{ls} \quad (20)$$

$$T'_{d0} = \frac{X_f + X_{md}}{R_f \omega_e} \quad (21)$$

$$T''_{d0} = \frac{X_{1d} + X_d{}'}{R_{1d}\omega_e} \quad (22)$$

$$T'_{q0} = \frac{X_{1q} + X_{mq}}{R_{1q}\omega_e} \quad (23)$$

$$T''_{q0} = \frac{X_{2q} + X_{mq}}{R_{2q}\omega_e} \quad (24)$$

$$T'_d = T'_{d0}\left(\frac{X'_d}{X_d}\right) \quad (25)$$

$$T''_d = T''_{d0}\left(\frac{X''_d}{X'_d}\right) \quad (26)$$

$$T'_q = \frac{X_{2q}}{R_{2q}} \quad (27)$$

$$T''_q = T''_{q0}\left(\frac{X''_q}{X'_q}\right) \quad (28)$$

Equations (29) to (38) represent the flux linkage-based Kirchhoff's voltage equations for the windings in d-q reference frame representation. All the previously used quantities are now expressed in per unit *. [9]

$$p\check{\psi}_d = -\omega\check{\psi}_q - \omega_B\check{R}_a\check{\iota}_d - \omega_B\check{v}_d \tag{29}$$

$$p\check{\psi}_q = \omega\check{\psi}_d - \omega_B\check{R}_a\check{\iota}_q - \omega_B\check{v}_q \tag{30}$$

$$p\check{\psi}_0 = -\omega_B\check{R}_a\check{\iota}_0 - \omega_B\check{v}_0 \tag{31}$$

$$p\check{\psi}_f = \frac{1}{T_d'}\left[-\check{\psi}_f + \check{\psi}_d + \left(\frac{\check{x}_d'}{(\check{x}_d - \check{x}_d')}\right)\check{E}_f\right] \tag{32}$$

$$p\check{\psi}_{1d} = \frac{1}{T_d''}\left[-\check{\psi}_{1d} + \check{\psi}_d\right] \tag{33}$$

$$p\check{\psi}_{1q} = \frac{1}{T_q'}\left(-\check{\psi}_{1q} + \check{\psi}_q\right) \tag{34}$$

$$p\check{\psi}_{2q} = \frac{1}{T_q''}\left(-\check{\psi}_{2q} + \check{\psi}_q\right) \tag{35}$$

$$\check{E}_f = \frac{\check{x}_{md}}{\check{R}_f}\check{V}_f \tag{36}$$

$$\check{\iota}_d = \frac{1}{\check{x}_d}\check{\psi}_d + \left(\frac{1}{\check{x}_d} - \frac{1}{\check{x}_d'}\right)\check{\psi}_f + \left(\frac{1}{\check{x}'_d} - \frac{1}{\check{x}''_d}\right)\check{\psi}_{1d} \tag{37}$$

$$\check{\iota}_q = \frac{1}{\check{x}_q}\check{\psi}_q + \left(\frac{1}{\check{x}_q} - \frac{1}{\check{x}'_q}\right)\check{\psi}_{1q} + \left(\frac{1}{\check{x}'_q} - \frac{1}{\check{x}''_q}\right)\check{\psi}_{2q} \tag{38}$$

On further arranging the equation in matrix – vector form, the state space model in terms of flux linkages is obtained, with a state equation (39) and an output equation (45). The complete formulation is shown using equations (39) to (46) with (47) representing the generated electromagnetic torque. [9]

$$\begin{bmatrix} \dot{\check{\psi}}_d \\ \dot{\check{\psi}}_q \\ \dot{\check{\psi}}_f \\ \dot{\check{\psi}}_{1d} \\ \dot{\check{\psi}}_{1q} \\ \dot{\check{\psi}}_{2q} \end{bmatrix} = \begin{bmatrix} 0 & -\omega & 0 & 0 & 0 & 0 \\ \omega & 0 & 0 & 0 & 0 & 0 \\ \frac{1}{T_d'} & 0 & -\frac{1}{T_d'} & 0 & 0 & 0 \\ \frac{1}{T_d''} & 0 & 0 & \frac{-1}{T_d''} & 0 & 0 \\ 0 & \frac{1}{T_q'} & 0 & 0 & \frac{-1}{T_q'} & 0 \\ 0 & \frac{1}{T_q''} & 0 & 0 & 0 & \frac{-1}{T_q'} \end{bmatrix} \begin{bmatrix} \check{\psi}_d \\ \check{\psi}_q \\ \check{\psi}_f \\ \check{\psi}_{1d} \\ \check{\psi}_{1q} \\ \check{\psi}_{2q} \end{bmatrix} + \begin{bmatrix} -(\check{R}_a + \check{R}_L) & 0 \\ 0 & -(\check{R}_a + \check{R}_L) \end{bmatrix} \begin{bmatrix} \check{\iota}_d \\ \check{\iota}_q \end{bmatrix} + \begin{bmatrix} 0 \\ 0 \\ \frac{1}{T_d'}\left(\frac{\check{X}'_d}{\check{X}_d - \check{X}'_d}\right)\left(\frac{\check{X}_{md}}{\check{R}_f}\right) \\ 0 \\ 0 \\ 0 \end{bmatrix} \check{E}_f \tag{39}$$

$$[\dot{X}] = [A_1][X] + [R][i] + [B]\check{E}_f \tag{40}$$

Where

$$[X] = \begin{bmatrix} \check{\psi}_d \\ \check{\psi}_q \\ \check{\psi}_f \\ \check{\psi}_{1d} \\ \check{\psi}_{1q} \\ \check{\psi}_{2q} \end{bmatrix}; [A_1] = \begin{bmatrix} 0 & -\omega & 0 & 0 & 0 & 0 \\ \omega & 0 & 0 & 0 & 0 & 0 \\ \frac{1}{T_d'} & 0 & -\frac{1}{T_d'} & 0 & 0 & 0 \\ \frac{1}{T_d''} & 0 & 0 & \frac{-1}{T_d''} & 0 & 0 \\ 0 & \frac{1}{T_q'} & 0 & 0 & \frac{-1}{T_q'} & 0 \\ 0 & \frac{1}{T_q''} & 0 & 0 & 0 & \frac{-1}{T_q'} \end{bmatrix}; [R] = \begin{bmatrix} -(\check{R}_a + \check{R}_L) & 0 \\ 0 & -(\check{R}_a + \check{R}_L) \end{bmatrix};\ B = \begin{bmatrix} 0 \\ 0 \\ \frac{1}{T_d'}\left(\frac{\check{X}'_d}{\check{X}_d - \check{X}'_d}\right)\left(\frac{\check{X}_{md}}{\check{R}_f}\right) \\ 0 \\ 0 \\ 0 \end{bmatrix} \tag{41}$$

$$[\dot{X}] = \{[A_1] + [A_2][A_3]\}[X] + [B]\check{E}_f \tag{42}$$

$$A_2 = \begin{bmatrix} -(\check{R}_a + \check{R}_L)\omega_B & 0 \\ 0 & -(\check{R}_a + \check{R}_L)\omega_B \\ 0 & 0 \\ 0 & 0 \\ 0 & 0 \\ 0 & 0 \end{bmatrix};\ A_3 = \begin{bmatrix} \frac{1}{\check{x}_d''} & 0 & \left(\frac{1}{\check{x}_d} - \frac{1}{\check{x}_d'}\right) & \left(\frac{1}{\check{x}_d'} - \frac{1}{\check{x}_d''}\right) & 0 & 0 \\ 0 & \frac{1}{\check{x}_q''} & 0 & 0 & \left(\frac{1}{\check{x}_q} - \frac{1}{\check{x}_q'}\right) & \left(\frac{1}{\check{x}_q'} - \frac{1}{\check{x}_q''}\right) \end{bmatrix}; \tag{43}$$

Further output voltage equation for the generator mode is considered as

$$\begin{bmatrix} \check{v}_d \\ \check{v}_q \end{bmatrix} = \begin{bmatrix} -\check{R}_L & 0 \\ 0 & -\check{R}_L \end{bmatrix} \begin{bmatrix} \check{\iota}_d \\ \check{\iota}_q \end{bmatrix} = \begin{bmatrix} -\check{R}_L & 0 \\ 0 & -\check{R}_L \end{bmatrix} \begin{bmatrix} \frac{1}{\check{x}_d''} & 0 & \left(\frac{1}{\check{x}_d} - \frac{1}{\check{x}_d'}\right) & \left(\frac{1}{\check{x}_d'} - \frac{1}{\check{x}_d''}\right) & 0 & 0 \\ 0 & \frac{1}{\check{x}_q''} & 0 & 0 & \left(\frac{1}{\check{x}_q} - \frac{1}{\check{x}_q'}\right) & \left(\frac{1}{\check{x}_q'} - \frac{1}{\check{x}_q''}\right) \end{bmatrix} \begin{bmatrix} \check{\psi}_d \\ \check{\psi}_q \\ \check{\psi}_f \\ \check{\psi}_{1d} \\ \check{\psi}_{1q} \\ \check{\psi}_{2q} \end{bmatrix} \tag{44}$$

This can be re-written in matrix form as follows

$$[Y] = [C][X] + [D]\breve{E}_f \tag{45}$$

Where $[Y] = \begin{bmatrix} \breve{v}_d \\ \breve{v}_q \end{bmatrix}$; $C = \begin{bmatrix} -R_L & 0 \\ 0 & -R_L \end{bmatrix} * [A_3]$; $D = [0]$

After calculating state matrix, the electromagnetic torque can be obtained from the solution of current matrix and the state matrix as shown,

$$\begin{bmatrix} \breve{i}_d \\ \breve{i}_q \end{bmatrix} = \begin{bmatrix} \frac{1}{\breve{x}''_d} & 0 & \left(\frac{1}{\breve{x}_d} - \frac{1}{\breve{x}'_d}\right) & \left(\frac{1}{\breve{x}'_d} - \frac{1}{\breve{x}''_d}\right) & 0 & 0 \\ 0 & \frac{1}{\breve{x}''_q} & 0 & 0 & \left(\frac{1}{\breve{x}_q} - \frac{1}{\breve{x}'_q}\right) & \left(\frac{1}{\breve{x}'_q} - \frac{1}{\breve{x}''_q}\right) \end{bmatrix} \begin{bmatrix} \breve{\psi}_d \\ \breve{\psi}_q \\ \breve{\psi}_f \\ \breve{\psi}_{1d} \\ \breve{\psi}_{1q} \\ \breve{\psi}_{2q} \end{bmatrix} \tag{46}$$

$$T_e = \{\breve{\psi}_d \breve{i}_q - \breve{\psi}_q \breve{i}_d\} \tag{47}$$

## 3. Proposed Magnetic Saturation Model

Dynamic performance of synchronous machine is closely associated with the effect of magnetic saturation of the iron core. However, an exact analysis of saturation can be very complex and impractical to use. It is very clear that an exact transient and steady-state mathematical model allow the saturation in both d-axis and the q-axis. For this, a complete magnetic field distribution inside the machine at every instant must be taken into consideration. However, it is computationally slow and expensive. Hence an assumption is considered for a salient pole generator along with assumption stated in section 2 i.e., air-gap along the q-axis is substantially more hence it is assumed that the impact of saturation is negligible for q-axis mutual inductances.

### 3.1 d-axis saturation

At rated speed, neglecting the armature reaction, per unit stator d-axis flux linkage is same as the per unit terminal voltage. This can be shown using equations (48-51). The unsaturated per unit flux linkage ($\psi_{t(unsat)}$)or per unit terminal voltage is directly proportional to the field current which is represented by the air-gap line as shown in Fig 5. But as saturation sets in, the relation is no longer linear, hence the saturated $\psi_{t(sat)}$ can be calculated from the proposed modified version of the Froelich's equation and the air gap line equation. Froelich's equation for any electrical machine defines its BH curve for the non-linear region using relation (55). Computation of $B$ & $H$ requires the hysteresis curve. For per unit analysis magnetic field $B$ can be modified as $\psi_t$ and magnetic field intensity $H$ as $I_f$. Hence the equation is further modified as (56). The constant $a$ and $b$ are calculated from OCC of the generator and the relations defined in (57).

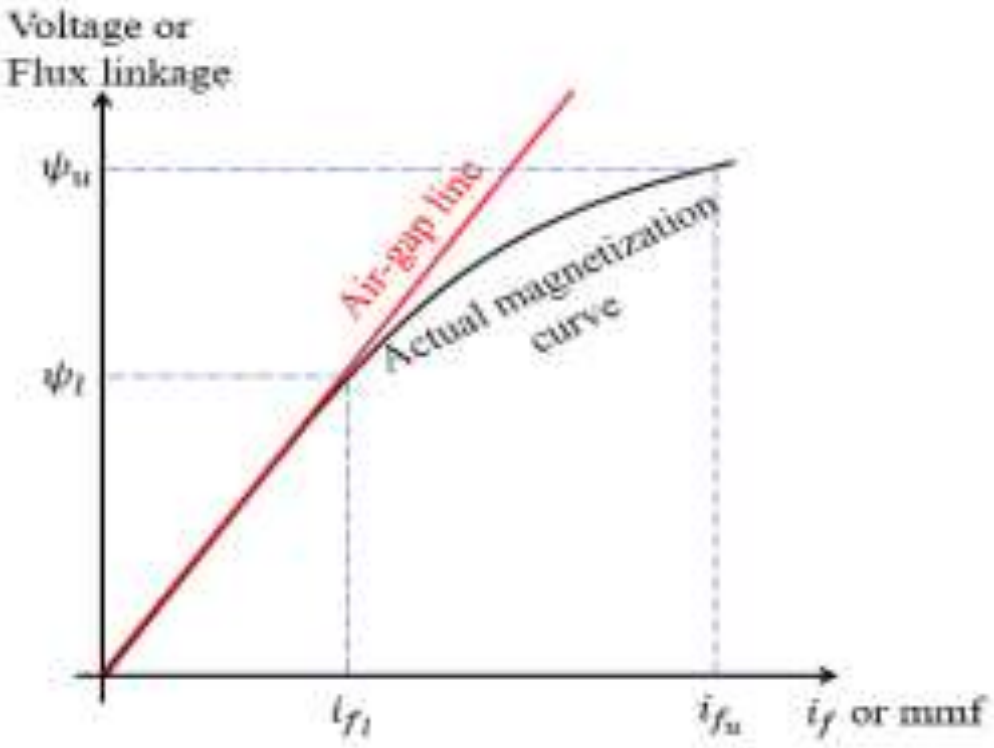

**Fig. 5.** *Hypothetical Magnetization Curve*

$$V_t = \omega\psi_d = \psi_d \tag{48}$$

$$\psi_d = x_{md} * i_f \tag{49}$$

$$\psi_q = 0 \tag{50}$$

$$\psi_t = \sqrt{\psi_d^2 + \psi_q^2} = \psi_d = V_t. \tag{51}$$

$$\frac{\psi_{t(unsat)}}{i_f} = x_{md(unsat)} \tag{52}$$

$$\frac{\psi_{t(sat)}}{i_f} = x_{md(sat)} \tag{53}$$

$$x_{md(sat)} = \frac{\psi_{t(sat)}}{\psi_{t(unsat)}} * x_{md(unsat)}. \tag{54}$$

$$B = \frac{H}{a+bH} \tag{55}$$

$$\psi_{t(sat)} = \frac{I_f}{a+bI_f} \tag{56}$$

$$a = \gamma - bI_{f_l},\ b = \frac{I_{f_u} - \gamma\psi_u}{\psi_u\left(I_{f_u} - I_{f_l}\right)},\ \gamma = \frac{I_{f_l}}{\psi_l}. \tag{57}$$

## 4. Verification of Saturation

For the validation of the proposed method, an experimental bench consisting of two machines coupled on same shaft was considered. Synchronous machine is the test machine and the other one is a prime mover as shown in Fig 6. figure. Synchronous generator with specifications shown in table 1 and 2 is considered and extracted the values of the constants $a$ *and* $b$ from its magnetization curve, $a = 0.48366$ and $b = 0.3478$ .From Fig.6. it can be said that the estimated magnetization curve is in close approximation with the actual one without much computational efforts.

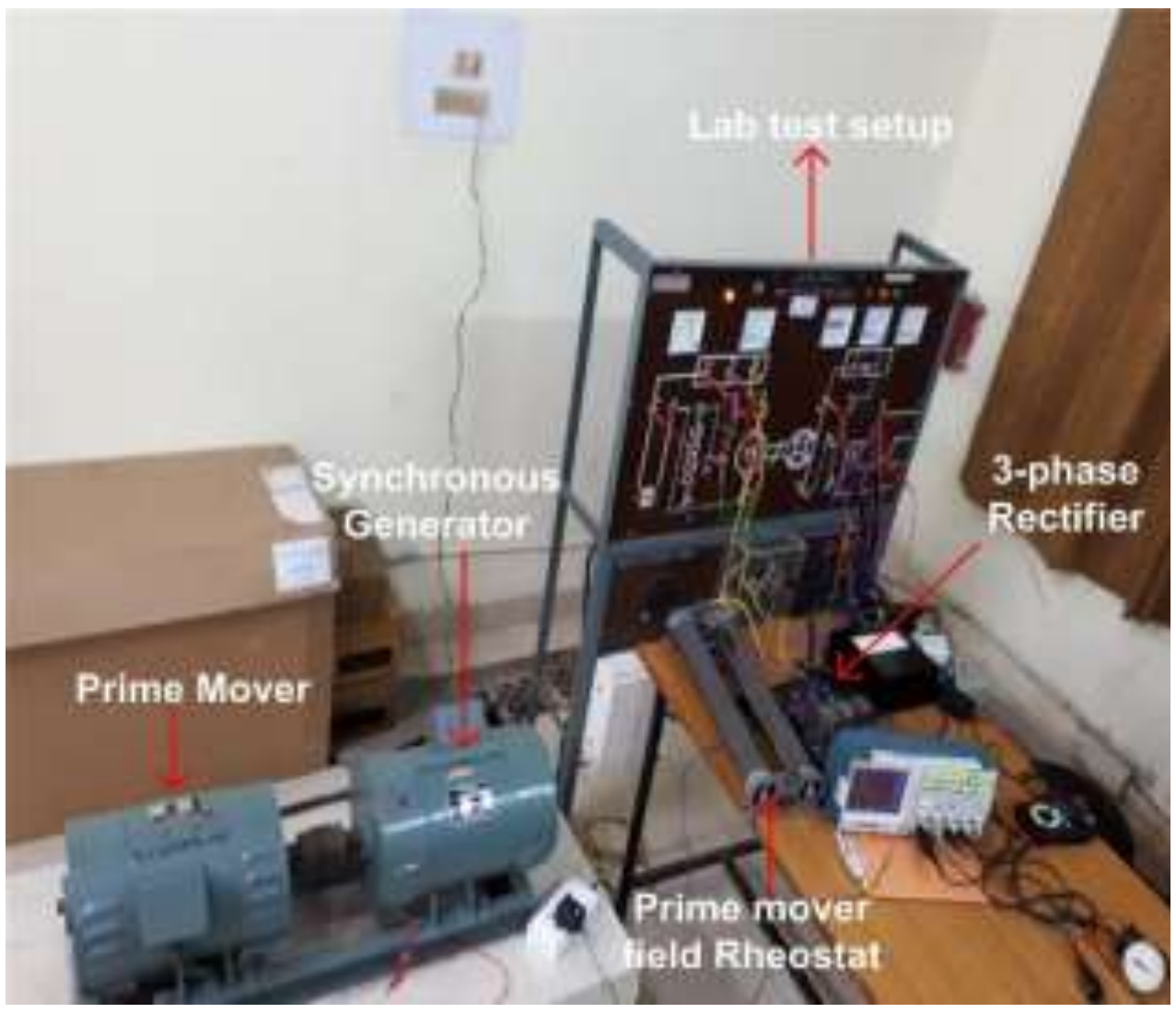

**Fig. 6.** *Experimental setup*

**Table 2** Machine Parameters

| Parameter | Standard value | Value in per unit |
|---|---|---|
| $X_d$ | 98.0679 Ω | 1.7085 |
| $X_d'$ | 32.6032 Ω | 0.568 |
| $X_d''$ | 10.1598 Ω | 0.177 |
| $X_q$ | 56.42 Ω | 0.9831 |
| $X_q'$ | 45.92 Ω | 0.8 |
| $X_q''$ | 30.3072 Ω | 0.528 |
| $X_{ls}$ | 4.879 Ω | 0.085 |
| $H'$ | 3.5 MJ/MVA | 3.5 MJ/MVA |
| $R_a$ | 0.1722 Ω | 0.003 |
| $R_f'$ | 1.722 Ω | 0.03 |
| Speed | 1500rpm | 1 |
| Line Voltage | 415 V | 1 |
| $I_{a_{rated}}$ | 3.3 A | 1 |
| Field voltage & current | 220 V, 0.7 A | 1 |

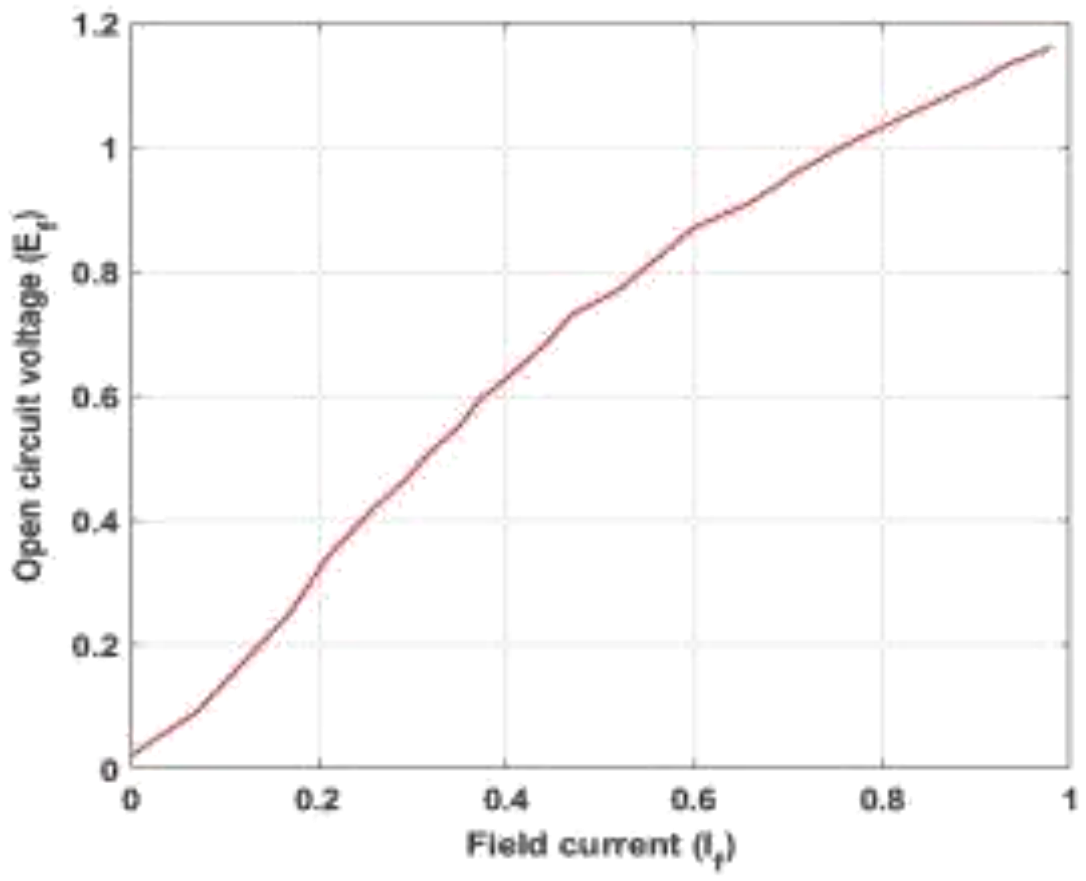

(a)

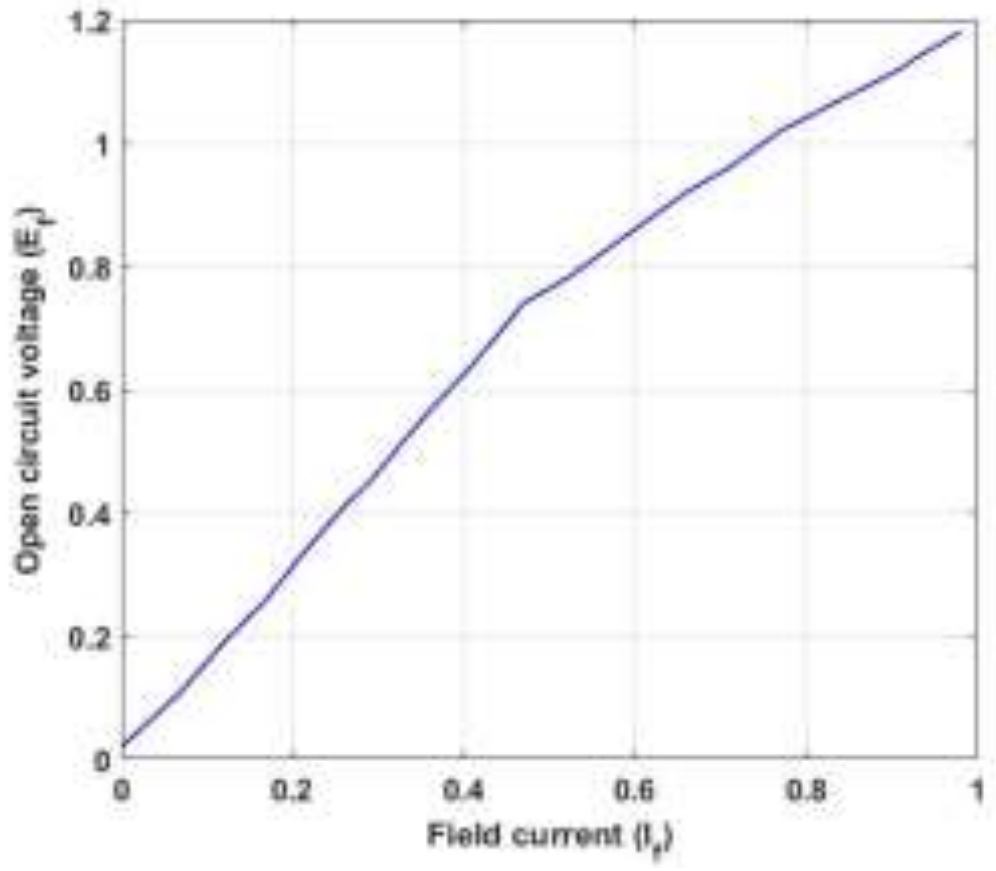

(b)

**Fig. 7.** *a) Actual magnetization curve from the experimental data* b) *Estimated magnetization curve from the proposed method*

## 5. Results and Discussion

### 5.1. Sudden short circuit test

Sudden short-circuit test is the sudden application of a three-phase short circuit across the machine. It represents a high load impact. This test helps us to test the transients due to damper windings. Fig.8.shows the output phase voltages, armature current and field current variations obtained from simulation and experiment respectively. These results by simulation and experimentation are approximately the same. The little difference is due to the assumptions considered in section 2.

### 5.2. Sudden open circuit test

In order to validate the proposed modelling, the machine is tested under sudden open circuit test following the short circuit test which was discussed in previous section, i.e., during steady state, an open circuit is performed on its three phases and then armature voltage, armature current and field currents are observed. Fig.8. shows the simulated and experimental waveforms before and after the sudden open circuit. From these figures it is observed that proposed model is in close agreement with the experimental results. Simulated and experimental results are in very similar shape.

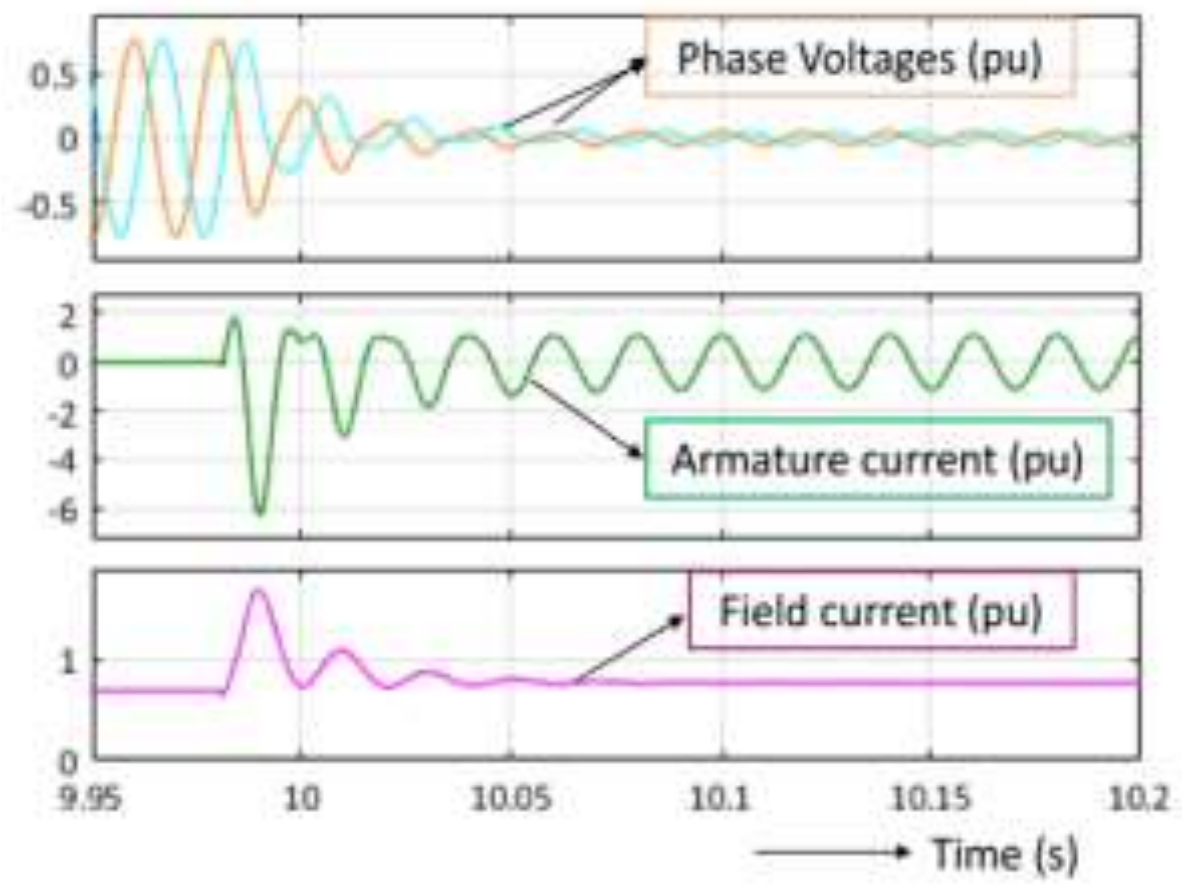

(a)

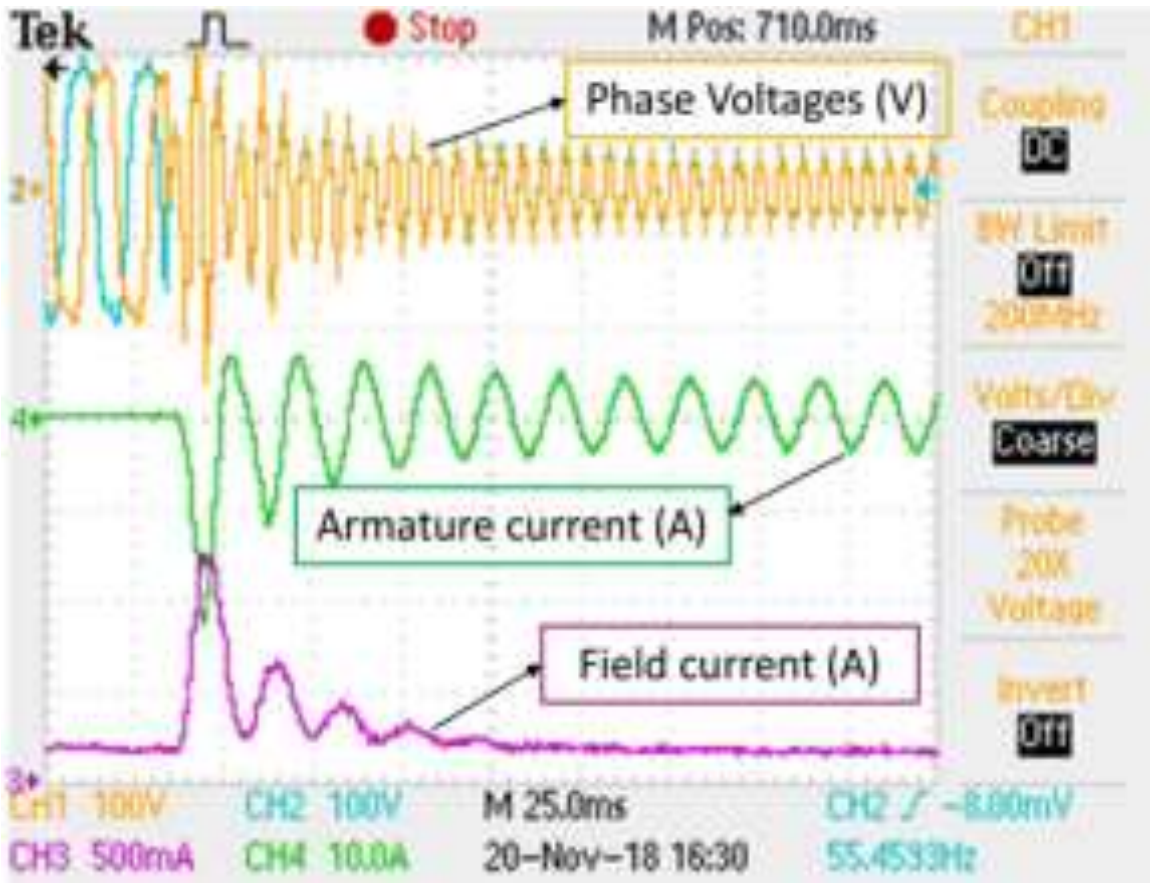

(b)

**Fig. 8.** *a) Simulated results of separately excited synchronous generator under sudden short circuit test*

*b) Experimental results of separately excited synchronous generator under sudden short circuit test*

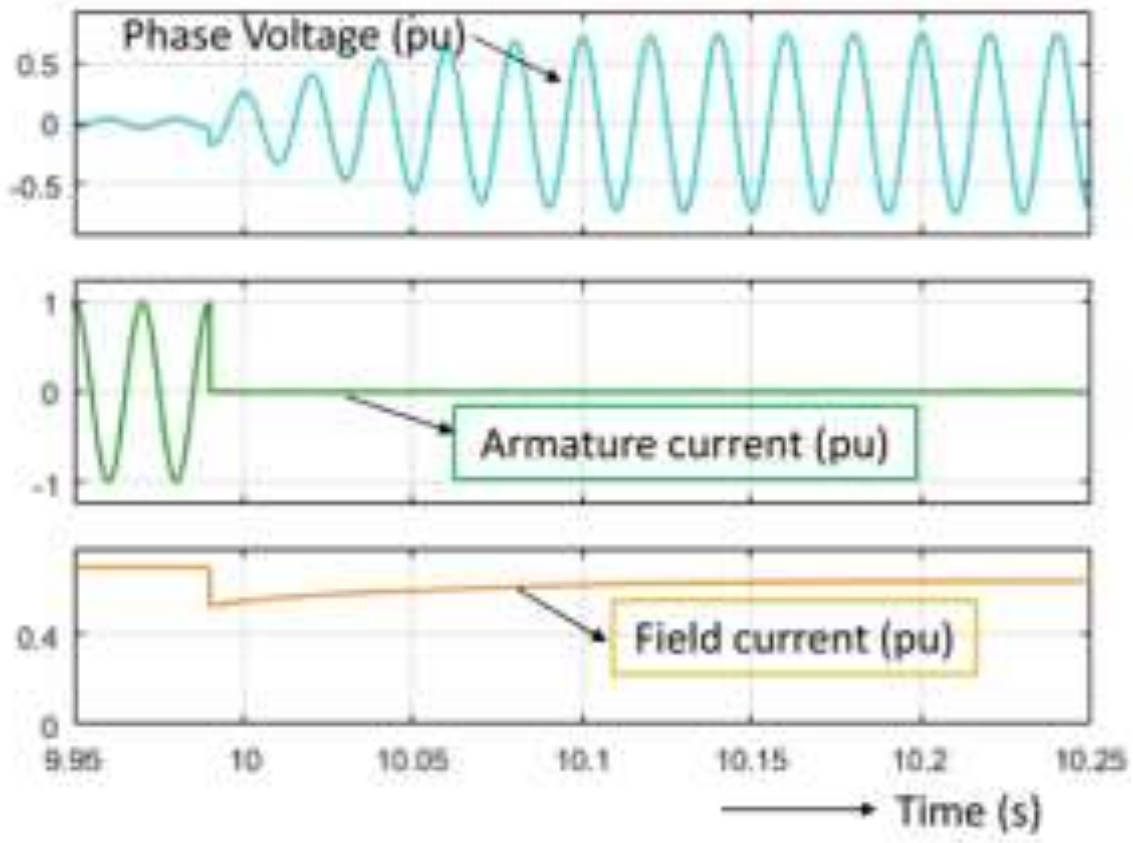

(a)

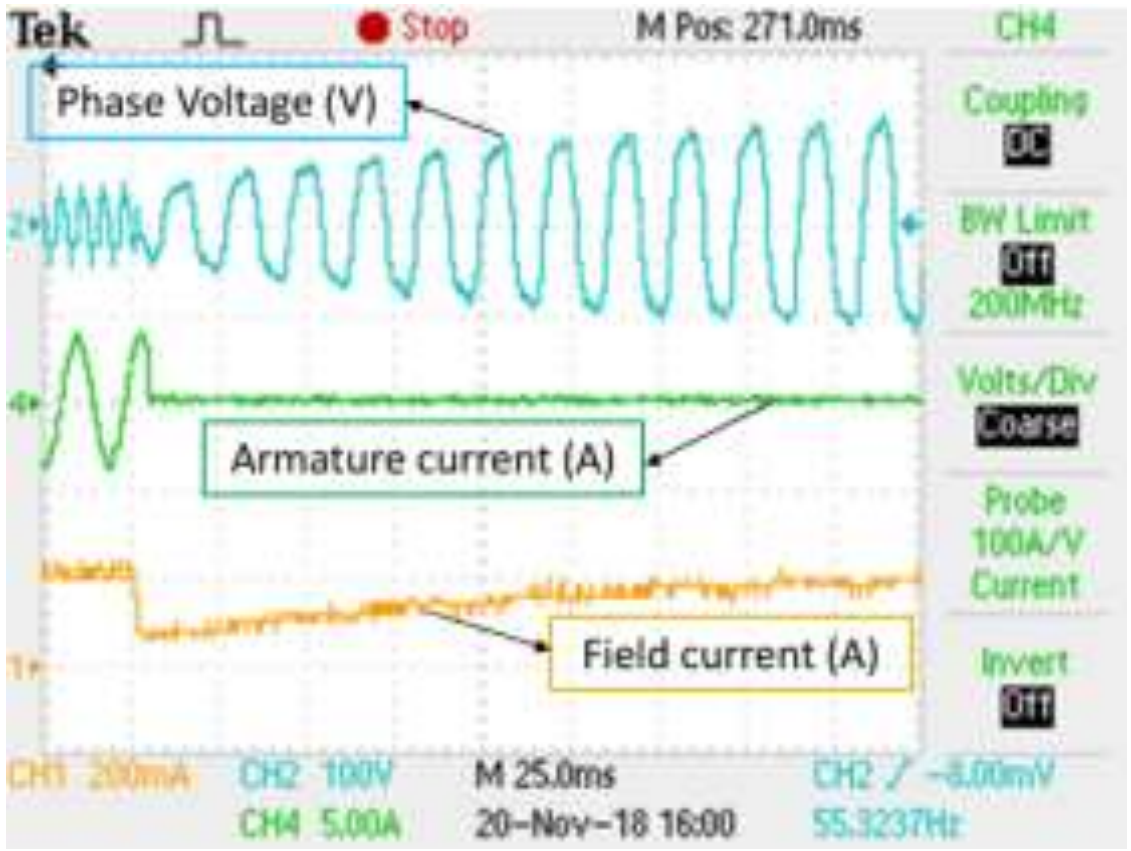

(b)

**Fig. 9.** *a) Simulated results of separately excited synchronous generator under sudden open circuit test*

*b) Experimental results of separately excited synchronous generator under sudden open circuit test*

### 5.3. Self-excitation

After successful validation of the machine model on separately excited synchronous generator, proposed saturation modelling is applied to self-excited synchronous generator in which magnetic saturation of the core plays a vital role. Figure shows the voltage buildup of a self-excited synchronous generator simulated results of the proposed scheme and corresponding experimental results of the same is also shown in Fig 10 (b). Along-side the variations in field current is also is compared.

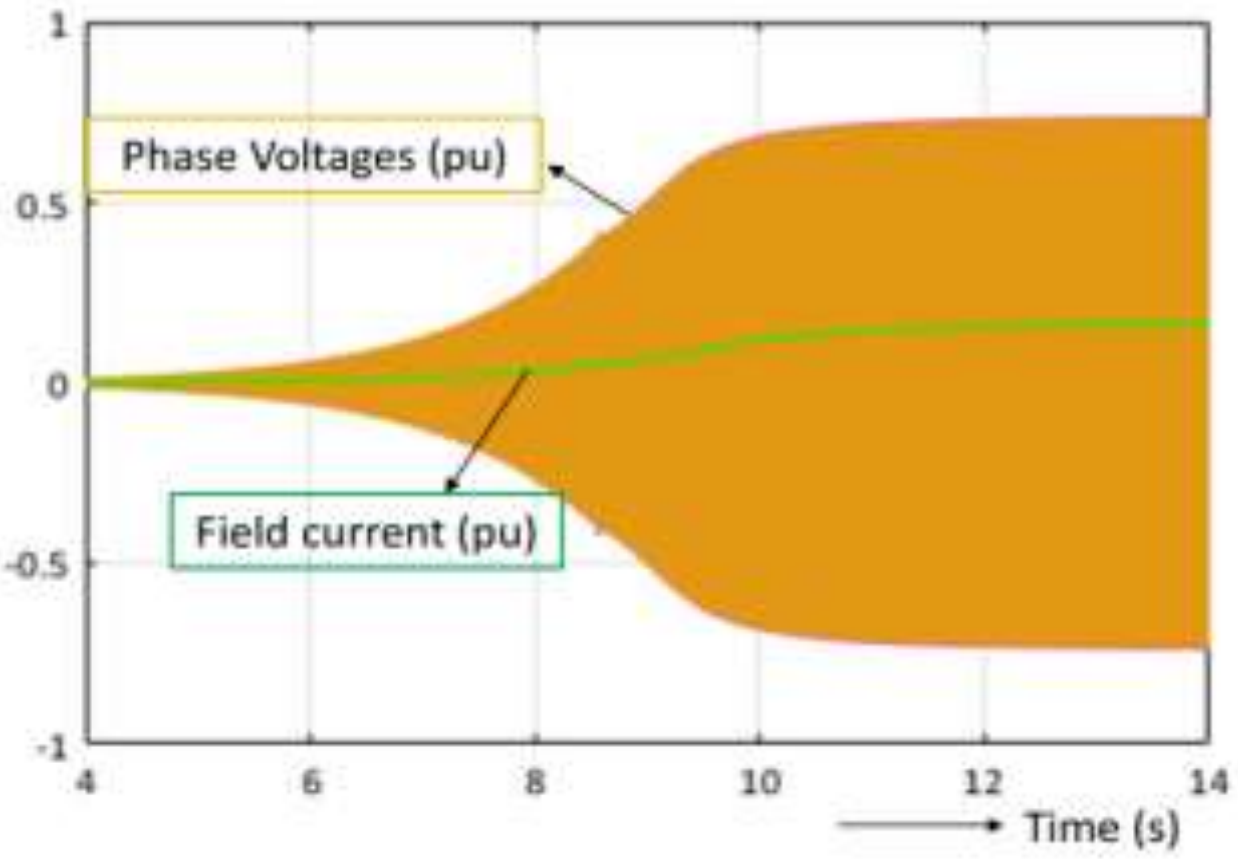

(a)

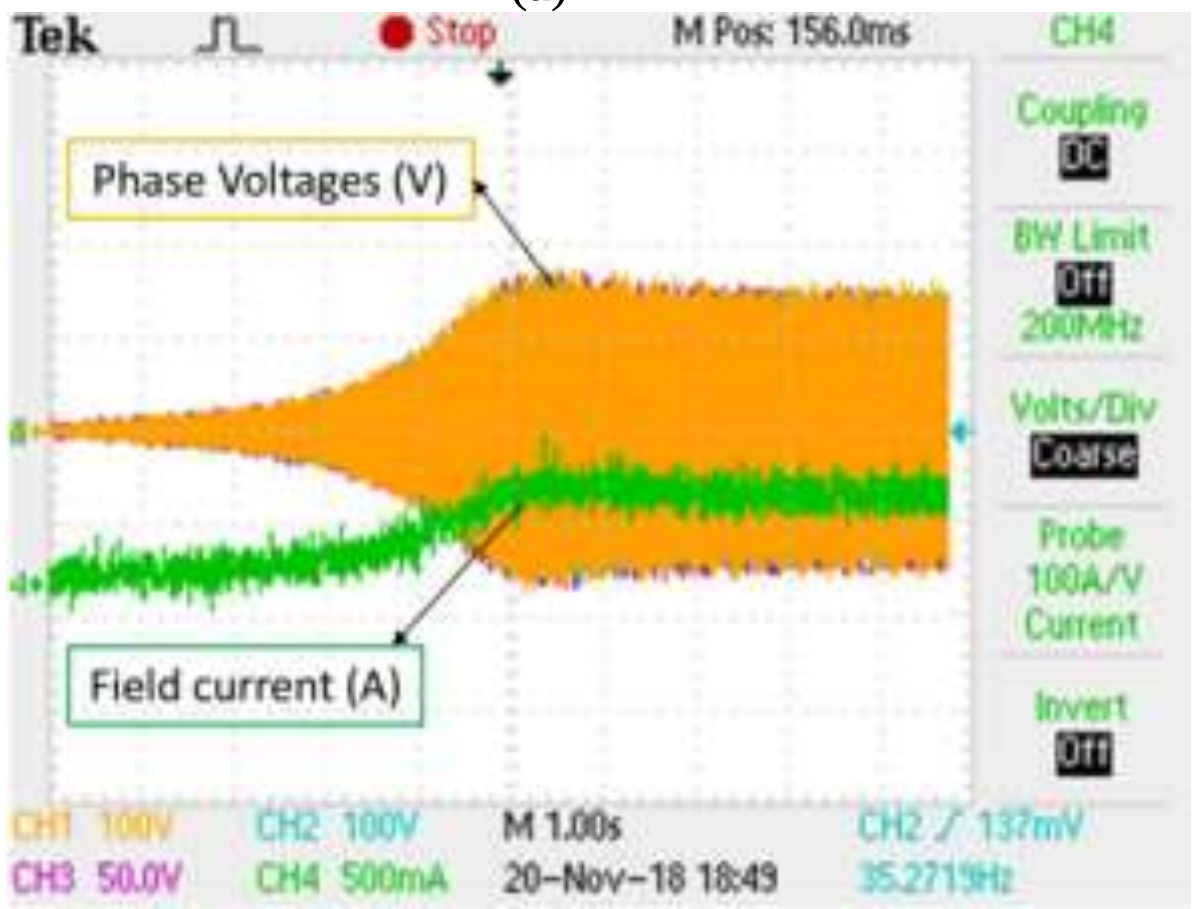

(b)

**Fig. 10**. *a) Simulated results of self-excited synchronous generator under voltage build-up*

*b) Experimental results of self-excited synchronous generator under voltage build-up*

## 6. Conclusion

In this paper, a complete modelling of synchronous generator taking in account of the existence of dampers and magnetic saturation, using only reactances and time constants as parameters, is presented without using tedious and complex computations. The Park's framework is used and the modelling is performed with the state space modelling. Saturation modelling procedure, described in the paper requires the knowledge of the machine's open circuit d-axis magnetising curve**.** The proposed method is simpler than those found in literature towards modelling the synchronous generator with magnetic saturation. The proposed model is validated by the experimental results under different system dynamics by applying sudden short circuit and sudden open circuiting the stator terminals by considering the self-excited synchronous generator where effect of magnetic saturation has a vital role.

**Table 3** Nomenclature

| | |
|---|---|
| $V_a, V_b, V_c, V_f, V_{1d}, V_{1q}, V_{2q}$ | Voltages applied in stator windings 'a', 'b', 'c', field winding 'f' and damper windings '1d', '1q', '2q'. |
| $i_a, i_b, i_c, i_f, i_{1d}, i_{1q}, i_{2q}$ | Currents through stator windings and field winding 'f' and damper windings '1d', '1q', '2q'. |
| $\psi_a, \psi_b, \psi_c, \psi_f, \psi_{1d}, \psi_{1q}, \psi_{2q}$ | Flux linkages with stator windings 'a', 'b', 'c', field winding 'f' and damper windings '1d', '1q', '2q'. |
| $V_d, V_q, V_0$ | Voltages applied across 'd', 'q','0' windings |
| $i_d, i_q, i_0$ | Currents passing in 'd', 'q','0' windings |
| $\psi_d, \psi_q, \psi_0,$ | Flux linkages with 'd', 'q','0' windings |
| $\psi_s^{3x1}, \psi_r^{4x1}$ | Stator flux linkage and rotor flux linkage vectors respectively |
| $i_s^{3x1}, i_r^{4x1}$ | Stator current and rotor current vectors respectively |
| $R_a, R_f, R_{1d}, R_{1q}, R_{2q}$ | Resistances of armature, field and dampers windings respectively |
| $L_a, L_b, L_c, L_f, L_{1d}, L_{1q}, L_{2q}$ | Self-Inductances of armature, field and damper windings |
| $M_{xy}$ | Mutual Inductance between various windings considered in a-b-c representation where $x = y =$ 'a', 'b', 'c','f','1d','1q','2q' |
| $L_{ss}^{3x3}, L_{sr}^{3x4}, L_{rs}^{4x3}, L_{rr}^{4x4}$ | Inductance matrices between stator windings, stator and rotor windings and rotor windings themselves. |
| $L_d, L_q, L_0$ | Self-Inductances of 'd', 'q','0' windings in d-q representation |
| $M_{ab}$ | Mutual Inductance between various windings considered in d-q representation where $a = b =$ 'd', 'q', 'f','1d','1q','2q' |
| $X_{ls}$ | Leakage reactance |
| $X_{md}, X_{mq}$ | d-axis and q-axis mutual reactances |
| $X_f, X_{1q}, X_{2q}, X_{1d},$ | Reactances for field and damper windings |
| $X_d, X_d', X_d''$ | d-axis steady state, transient and sub transient reactances |
| $X_q, X_q', X_q''$ | q-axis steady state, transient and sub transient reactances |
| $T_{do}', T_{do}'', T_{qo}', T_{qo}''$ | d-axis and q-axis transient and sub transient open circuit time constants respectively. |
| $T_d', T_d'', T_q', T_q''$ | d-axis and q-axis transient and sub transient short circuit time constants respectively |
| $E_f$ | Excitation emf |
| $V_f$ | Applied field winding voltage |
| $R_L$ | Load Resistance |
| $\omega$ | Synchronous frequency |
| $\omega_B$ | Base frequency |
| $[A_1] + [A_2][A_3]$ | State Matrix |
| $[B]$ | Input Matrix |
| $[C]$ | Output matrix |
| $T_e$ | Electromagnetic torque |
| $\mathrm{x}_{\mathrm{md(unsat)}}$ | Unsaturated value of d-axis mutual reactance |
| $\mathrm{x}_{\mathrm{md(sat)}}$ | Saturated value of d-axis mutual reactance |
| $\psi_{\mathrm{t(sat)}}$ | Saturated value of net flux linkage |
| $\psi_{\mathrm{t(unsat)}}$ | Unsaturated value of net flux linkage |
| $\mathrm{I}_{\mathrm{f_l}}$ | Lower limit of field current from OCC |
| $\mathrm{I}_{\mathrm{f_u}}$ | Upper limit of field current from OCC |
| $\psi_{\mathrm{u}}$ | Upper limit of net flux linkage from OCC |
| $\psi_{\mathrm{l}}$ | Lower limit of net flux linkage from OCC |
| $H'$ | Inertia constant of generator |

*****Quantities with '~' sign represent the per unit versions of the actual quantities